\documentclass[10pt, conference]{IEEEtran}
\IEEEoverridecommandlockouts
\usepackage{cite}
\usepackage{amsmath,amssymb,amsfonts}
\usepackage{algorithmic}
\usepackage{graphicx}
\usepackage{textcomp}
\usepackage{makecell}
\usepackage{booktabs} 
\usepackage{amsmath} 
\usepackage{amsfonts} 
\usepackage{algorithm}
\usepackage{graphicx} 
\usepackage{array} 
\usepackage{algorithmic}
\usepackage{subcaption}
\usepackage{graphicx}
\usepackage{textcomp}
\usepackage[table]{xcolor}
\usepackage{rotating} 
\usepackage{multirow}
\usepackage{tabularx}
\usepackage{diagbox}
\usepackage{caption}
\usepackage{footnote}
\usepackage{enumitem}
\usepackage{hyperref}
\usepackage{makecell}
\usepackage{comment}
\usepackage{amssymb} 

\def\BibTeX{{\rm B\kern-.05em{\sc i\kern-.025em b}\kern-.08em
    T\kern-.1667em\lower.7ex\hbox{E}\kern-.125emX}}
\begin{document}

\newcommand{\rqin}[1]{\textcolor{orange}{#1}}
\newcommand{\jx}[1]{\textcolor{blue}{JX: #1}}
\newcommand{\dcheng}[1]{\textcolor{red}{DC: #1}}
\newcommand{\amir}[1]{\textcolor{orange}{ #1}}
\title{Tiny-Align: Bridging Automatic Speech Recognition and Large Language Model on Edge}

\author{Authors \\ Institutes
}
\author{Ruiyang Qin$^{1, 2, 3}$, Dancheng Liu$^{3}$, Gelei Xu$^{2}$, Amir Nassereldine$^{3}$, Zheyu Yan$^{2}$, Chenhui Xu$^{3}$, Yuting Hu$^{3}$, \\ Shaocong Wang$^{2}$, X. Sharon Hu$^{2}$, Jinjun Xiong$^{3}$ and Yiyu Shi$^{2}$
\\ $^{1}$Villanova University, $^{2}$University of Notre Dame, $^{3}$University at Buffalo--SUNY}


\maketitle

\begin{abstract}
The combination of Large Language Models (LLM) and Automatic Speech Recognition (ASR), when deployed on edge devices (called edge ASR-LLM), can serve as a powerful personalized assistant to enable audio-based interaction for users. Compared to text-based interaction, edge ASR-LLM allows accessible and natural audio interactions. Unfortunately, existing ASR-LLM models are mainly trained in high-performance computing environments and produce substantial model weights, making them difficult to deploy on edge devices. More importantly, to better serve users' personalized needs, the ASR-LLM must be able to learn from each distinct user, given that audio input often contains highly personalized characteristics that necessitate personalized on-device training. Since individually fine-tuning the ASR or LLM often leads to suboptimal results due to modality-specific limitations, end-to-end training ensures seamless integration of audio features and language understanding (cross-modal alignment), ultimately enabling a more personalized and efficient adaptation on edge devices. However, due to the complex training requirements and substantial computational demands of existing approaches, cross-modal alignment between ASR audio and LLM can be challenging on edge devices. In this work, we propose a resource-efficient cross-modal alignment framework that bridges ASR and LLMs on edge devices to handle personalized audio input. Our framework enables efficient ASR-LLM alignment on resource-constrained devices like Raspberry Pi 5 (8GB RAM), achieving 50x training time speedup while improving the alignment quality by more than 50\%. To the best of our knowledge, this is the first work to study efficient ASR-LLM alignment on resource-constrained edge devices.

\end{abstract}

\section{Introduction}
\label{sec:intro}
The Large Language Model (LLM) deployed on the edge device (edge LLM) can serve as a personalized assistant, guaranteeing data privacy \cite{qin2023enabling} and continuous service without relying on stable internet connections \cite{sarabi2023llm, qin2024language, hu2025combating}. Edge LLMs normally interact with users based on text (text-based interaction). Yet, it is more natural and accessible for users to interact with edge LLMs, their personalized assistants, via audio \cite{nassereldine2024pi}. To enable the audio-based interaction,  the Automatic Speech Recognition (ASR) models can be employed \cite{yu2016automatic} to simply take the audio as the input and send the converted text into the edge LLM \cite{yang2024mala}. Existing works, on the other hand, have shown that this simple combination of ASR and LLM can fail in cases where the audio has no corresponding text \cite{bai2024seed}, or the mismatching pre-trained knowledge in the ASR and the edge LLM can compromise the performance of their combination.

\begin{figure}[t!]
  \centering
  \begin{subfigure}[b]{\linewidth}
    \centering
    \includegraphics[width=0.95\linewidth]{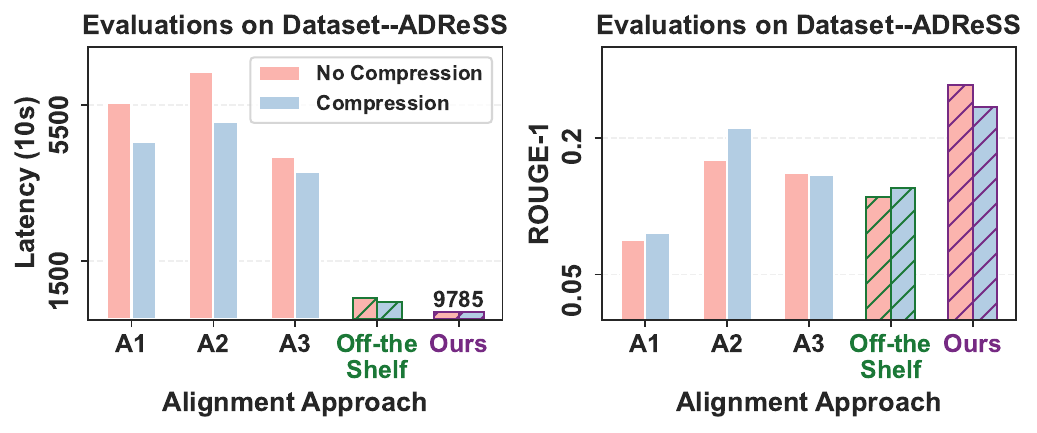}
  \end{subfigure}
  
  \hfill
  
  \begin{subfigure}[b]{\linewidth}
    \centering
    \includegraphics[trim=0 395 507 0, clip, width=0.95\linewidth]{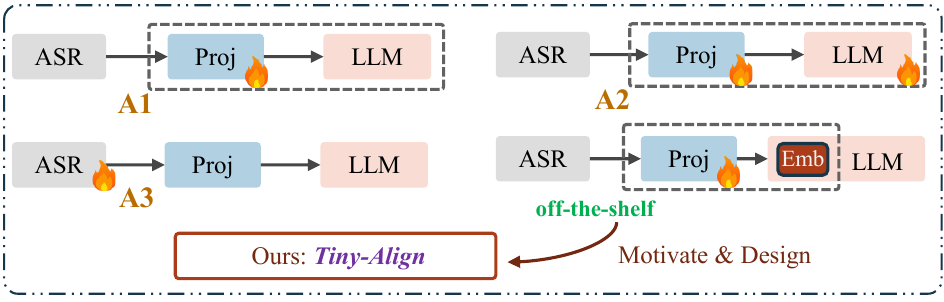}
  \end{subfigure}

  \caption{Evaluation of ASR-LLM alignment using Whisper-base and Gemma-2-2B with and without GPTQ (G) compression. Comparison includes three existing approaches (A1, A2, A3) and a preliminary design using only projector and LLM embedding (\textit{off-the-Shelf}). This preliminary result motivates our work (detailed in Fig.~\ref{fig:method}), which outperforms existing approaches.}

  \label{fig:preliminary_compare_method}
\end{figure}

Different from simply combining the ASR and the edge LLM, past works \cite{mittal2024salsa} concatenate them at the feature level via the projector. In this concatenation, the ASR functions as the audio encoder and produces the audio embedding, which will then be sent to the edge LLM for LLM generation (edge ASR-LLM). Usually by default, the projector is a simple multi-layer perception (MLP) that matches the dimensionality of the audio embedding with the LLM. While the feature-level concatenation can overcome the problems in the simple combination of the ASR and the edge LLM, it often requires end-to-end training so the audio embeddings can be properly understood by the edge LLM (cross-modal alignment).

For cross-modal alignment, there are three main approaches to end-to-end training, as shown right part of Fig.~\ref{fig:preliminary_compare_method}. The first approach (A1) freezes both the ASR encoder and LLM while training only the projector \cite{wu2023next}. This method updates the projector parameters by comparing LLM-generated outputs: one from projector-transformed audio embeddings and another from the corresponding text input. The second approach (A2) extends the first approach by jointly training \cite{ye2024x} both the projector and LLM, creating a larger learning space. The third approach (A3) takes a different direction by training only the ASR encoder \cite{zhu2023languagebind} while keeping the projector and LLM frozen.


The three approaches, while demonstrating decent performances in cross-modal alignment, can involve a huge size of parameter calculation upon the complex hyperparameters and the model tuning space. In this way, the huge resource budget can be prohibitive for cross-modal alignment on edge devices. Under the existing cross-modal alignment approaches, even if we choose the compressed model with smaller parameters, we can expect either significant performance downgrading due to the lowering of parameter size or the still prohibitively high resource usage due to the internal training process design. As shown left part of Fig.~\ref{fig:preliminary_compare_method}, we provide a pre-evaluation of the three alignment approaches on the concatenation of an ASR model--Whisper \cite{radford2023robust} and the edge LLM--Gemma-2-2B \cite{team2024gemma}. We also employ the compress Gemma-2-2B based on GPTQ \cite{frantar2022gptq} (G). The ASR and the edge LLM are connected by a two-layer MLP. In terms of training time (latency) and performance, we do not observe significant differences between the original model and its compressed version.

For approaches A1 and A2, the LLM generation involved can consume a huge amount of resources, leading to a longer convergence time. The approach of A2, which takes the projector and the LLM for training, outperforms the other two approaches but it also takes the most resources. The approach of  A1 is more lightweight, but it still uses the LLM, and its reduction in finetuning resources leads to significant performance degradation. Meanwhile, The A1's lower performance than the A2 can be due to the simple projector design. As shown in Fig.~\ref{fig:preliminary_compare_method}, while the A2 obtains decent performance, it also has the highest latency mostly due to the LLM generation. The \textit{off-the-shelf} is tentatively made to avoid the LLM generation. To do that,  we pick and use only its embedding layer to avoid the LLM generation. 
The embedding layer interfaces directly with the ASR encoder through a well-designed projector, which requires both sufficient embedding capacity and architectural compatibility with the ASR encoder and the rest components of the edge LLM.
In Fig.~\ref{fig:preliminary_compare_method}, we compare this design as \textit{off-the-shelf} with the three existing approaches. 
Surprisingly, we discovered that this design can outperform A1, A2, and even A3 in terms of performance and training latency.

Motivated by our observations, in this paper, we propose a resource-efficient alignment framework to enable cross-modal alignment between ASR and LLM on-edge devices, effectively balancing resource usage and end-to-end training performance. In our alignment framework, we design a novel projector that provides strong expressive power while consuming minimal training resources. Given its effectiveness in edge ASR-LLM scenarios, we name our framework as \textbf{Tiny-Align}. Our experiments demonstrate that Tiny-Align achieves significant performance improvements in aligning ASR encoders with various edge LLMs, requiring minimal time and computational resources. 

Our major contributions can be summarized as follows:

\begin{itemize}
\item We propose a novel approach (Tiny-Align) to enable the efficient alignment of ASR and LLM on resource-constrained edge devices. To the best of our knowledge, Tiny-Align is the first of its kind. Compared to existing alignment methods that require server-scale computing resources, Tiny-Align achieves training convergence 50$\times$ faster while
Improving the quality of the results under ROUGE-1 and ROUGE-L scores by more than 50\%.
\item  Tiny-Align employs a new projector design (BridgeFormer) based on a transformer encoder architecture but without the positional encoding component.  BridgeFormer provides a larger embedding space than existing MLPs or DNN-based designs and is shown to outperform the latter.
\item We further introduce a novel instruction injection mechanism for Tiny-Align’s inference, and show that it can further improve the quality of the results by about 50\% compared to Tiny-Align without instruction injection.
\end{itemize}
All our results have been conducted on datasets collected from individuals with various disabilities, such as dementia, aphasia, and speech-language impairments (SLI), hence exhibiting a wide range of speech needs. We show that Tiny-Align can support high-quality interactions for them through an ASR interface.


\begin{figure*}[h]
  \centering
  \includegraphics[trim=0 345 220 0, clip, width=1.\linewidth]{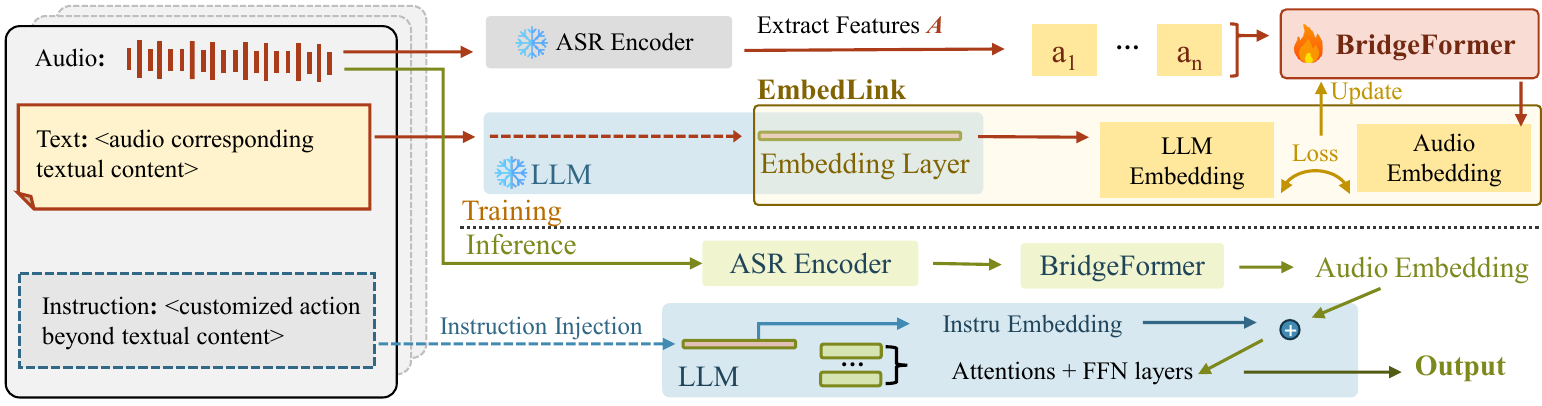}
  \caption{Tiny-Align Framework. It unifies the trainable parameters into the designed transformer-based projector (BridgeFormer), which is trained to minimize the loss between the audio embedding made by itself and the textual embedding made through the original LLM embedding layer.}
  \vspace{2ex}
  \label{fig:method}
\end{figure*}

\section{Background}
\subsection{Automatic Speech Recognition onto edge LLM} 
\label{background:ASR}

Extending edge LLM's capability to process audio input can significantly enhance user accessibility. ASR models serve as the bridge between audio and text modalities, with several prominent approaches offering different trade-offs between performance and resource efficiency.

Wav2vec2 \cite{baevski2020wav2vec} pioneered an efficient approach to audio processing by directly learning representations from raw waveforms. Through a multi-layer convolutional network, it extracts context-dependent representations that effectively capture speech patterns. The model employs contrastive learning to identify relevant features and discern patterns in the audio signal. 
While its streamlined architecture makes it particularly efficient for edge deployment, wav2vec2 has demonstrated robust performance in various scenarios \cite{barcovschi2023comparative}, including child speech recognition, highlighting its capability to effectively capture speech features across diverse conditions.

Whisper \cite{radford2023robust} and Conformer \cite{gulati2020conformer} represent another advancement in ASR technology, leveraging transformer architectures for improved accuracy. These models process audio through self-attention mechanisms to capture long-range dependencies in speech signals, achieving superior performance in complex transcription tasks across various languages and acoustic conditions \cite{liu2024automatic}. However, their sophisticated architecture requires more computational resources compared to wav2vec2.

More recent approaches like Diffsound \cite{yang2023diffsound}, AudioLM \cite{borsos2023audiolm}, and Tango2 \cite{majumder2024tango} adopt generative frameworks that extend beyond simple transcription. Based on diffusion models or autoregressive generation, these ASRs can not only recognize speech but also model the underlying audio distribution. While these models achieve state-of-the-art performance in complex scenarios, their resource-intensive nature makes them highly challenging to deploy in edge environments without significant optimization.

\subsection{Cross-modal Alignment on Edge}
One key challenge in personalizing edge LLM is adapting to users' preferred input modality while operating within the constraints of limited computational resources \cite{luo2023open, qin2024empirical, qin2024nvcim, qin2024robust}. Cross-modal alignment enables the integration of non-textual inputs by combining specialized machine learning models for various modalities with the edge LLM which provides advanced reasoning capabilities for analyzing cross-modal data \cite{qian2024linguistic}. However, this alignment process demands substantial resources, making deployment on edge devices particularly challenging \cite{xing2018enabling, qin2024fl}. For example, on resource-constrained devices like Raspberry Pi 5 (8GB RAM, quad-core Cortex-A76), cross-modal alignment can quickly exhaust available memory and computational resources, with processing latency becoming prohibitive for real-time applications. These constraints highlight the need for optimizing cross-modal alignment specifically for edge deployment.





\subsection{Edge ASR-LLM Alignment for People with Speech Difficulties}


While edge ASR-LLM alignment enhances accessibility for general users, it holds particular promise for individuals with speech difficulties. In this work, we demonstrate the effectiveness of our alignment approach using datasets from individuals with dementia, aphasia, and specific language impairments (SLI), showcasing how edge ASR-LLM can adapt to and understand users despite significant speech challenges. This capability becomes increasingly crucial given the growing population affected by these conditions—over 6.7 million people with dementia, 2 million with aphasia, and more than 180,000 new cases annually \cite{kramarow2024dementia}. For these individuals, traditional text-based interaction is infeasible due to motor and cognitive challenges, making voice-based interaction not merely an alternative, but a necessity for healthcare delivery.



\section{Proposed Work}


In this section, we first provide an overview of our Tiny-Align framework shown in Fig.~\ref{fig:method}. We then delve into the technical details, beginning with the pivotal component: the projector design. Following this, we explain how the projector collaborates with the LLM embedding layer. We then elaborate on the types of ASR models suitable for our edge ASR-LLM alignment framework. Finally, we introduce our instruction injection method, which enables projector-generated embeddings to better facilitate LLM generation.

\subsection{Framework Overview}
As shown in Fig.~\ref{fig:method}, Tiny-Align operates in two modes. During the training mode (upper half in Fig.~\ref{fig:method}), our proposed BridgeFormer establishes a shared embedding space through the proposed EmbedLink, enabling effective cross-modal alignment between ASR and LLM, guided by the LLM embedding layer. Additionally, we analyze and identify optimal ASR models that balance feature extraction capability with resource efficiency, ensuring Tiny-Align's effectiveness across different deployment scenarios. During the inference model (lower half in Fig.~\ref{fig:method}), ASR-extracted features are transformed via BridgeFormer into LLM-compatible embeddings, which can concatenate with instructions under our proposed instruction injection mechanism. Either the LLM-compatible embeddings or the concatenations will go through the LLM for generating the output.


\subsection{BridgeFormer: Transformer-based Projector}

Motivated by our preliminary study in Fig.~\ref{fig:preliminary_compare_method}, where a simple MLP projector demonstrates promising results in edge ASR-LLM alignment, we design an enhanced projector architecture that offers a larger audio embedding space where the embeddings can also be understood by the LLM, while maintaining resource efficiency. This design is driven by two key requirements: the ability to support continuous learning during user interactions and architectural compatibility with LLMs to ensure effective embedding transformation. As the sole trainable module in Tiny-Align, our projector aims to leverage the computational resources saved from avoiding LLM generation and training, allowing for a moderately larger architecture while remaining within edge device constraints.

To meet these design criteria, we introduce a transformer encoder architecture into the projector, named \textbf{BridgeFormer} to reflect its role in bridging ASR and LLM models. Compared to classical MLPs or deep neural networks, the transformer architecture offers several advantages: a larger embedding space with enhanced learning potential, natural alignment with LLM's architecture (as transformers were originally designed for natural language processing) \cite{vaswani2017attention, qin2021ibert}, and proven effectiveness in handling sequential data. Furthermore, since BridgeFormer processes ASR-extracted audio features where temporal positioning is already encoded, we remove the traditional positional encoding component, simplifying the architecture and improving its compatibility with both ASR and edge LLM.

\begin{figure}[t]
  \centering
  \includegraphics[trim=0 385 455 0, clip, width=1.\linewidth]{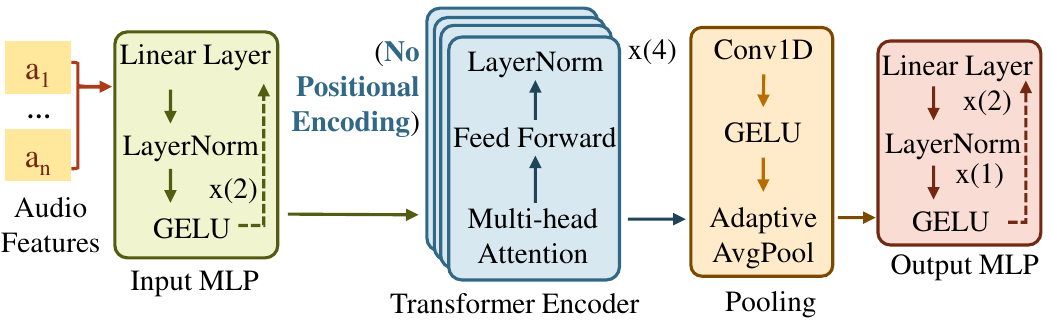}
  \caption{Architecture of BridgeFormer.}
  \label{fig:BridgeFormer}
\end{figure}


Consider audio features extracted from an ASR encoder with shape $[1, N, D_a]$, where $N$ represents the audio sequence length and $D_a$ is the ASR embedding dimension. BridgeFormer first employs an input MLP to project these features into a higher-dimensional space, creating initial embeddings of shape $[1, N, H]$, where $H$ is the hidden dimension. These embeddings then pass through the transformer encoder, where $M$ attention heads across $E$ encoder layers to process and refine the features while maintaining the sequence dimensions. To prevent overfitting and control sequence length, an adaptive pooling layer reshapes the sequence to a predefined token size $T$, producing outputs of shape $[1, T, H]$. Finally, an output MLP transforms these features to match the LLM embedding dimension $D_l$, resulting in LLM-compatible embeddings of shape $[1, T, D_l]$. This process can be formally expressed as:

\begin{equation}
   E_{\text{in}} = \text{MLP}_{\text{input}}(F_{\text{audio}}) \in \mathbb{R}^{1 \times N \times H}
\end{equation}
where $F_{\text{audio}} \in \mathbb{R}^{1 \times N \times D_a}$ represents the input audio features from ASR.

\begin{equation}
   E_{\text{trans}} = \text{Transformer}_{M,E}(E_{\text{in}}) \in \mathbb{R}^{1 \times N \times H}
\end{equation}
where $M$ denotes the number of attention heads and $E$ represents the number of encoder layers.

\begin{equation}
   E_{\text{pool}} = \text{Pool}(E_{\text{trans}}) \in \mathbb{R}^{1 \times T \times H}
\end{equation}
where $T$ is the target token size (discussed in the following section).

\begin{equation}
   E_{\text{out}} = \text{MLP}_{\text{output}}(E_{\text{pool}}) \in \mathbb{R}^{1 \times T \times D_l}
\end{equation}
where $D_l$ represents the LLM embedding dimension.

BridgeFormer's architecture is designed for flexibility and resource adaptability. The model can be easily scaled to match available edge resources through three key parameters: hidden dimension (controlling embedding capacity), number of attention heads (affecting feature relationship modeling), and number of transformer encoder layers (determining processing depth). As shown in Fig.~\ref{fig:BridgeFormer}, BridgeFormer comprises four essential components: (1) an input MLP that projects features into a high-dimensional embedding space, (2) transformer encoder layers that process and refine these embeddings, (3) a pooling layer that prevents overfitting and controls dimensionality, and (4) an output MLP that generates the final LLM-compatible embeddings.

\subsection{EmbedLink: Cross-modal Alignment via Embedding Layer}

With BridgeFormer's architecture established, we now focus on its training methodology for creating a shared embedding space. We propose EmbedLink, a simple yet effective pipeline that avoids resource-intensive LLM generation during cross-modal alignment. Our approach leverages a key insight: LLM's text processing fundamentally relies on embeddings from its embedding layer. Therefore, if BridgeFormer can transform audio features into embeddings that closely match those generated by the LLM's embedding layer from corresponding text input, we can achieve effective cross-modal alignment without the computational overhead of full LLM inference.

\begin{figure}[t]
  \centering
  \includegraphics[trim=0 320 615 0, clip, width=1.\linewidth]{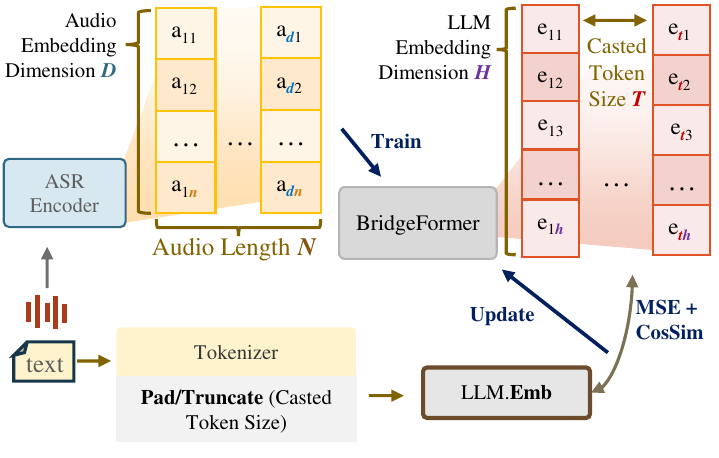}
  \caption{Demonstration of the EmbedLink usage in the BridgeFormer training. The EmbedLink uses casted token size, LLM embedding layer, and two loss functions to update parameters in BridageFormer.}
  \label{fig:projector}
\end{figure}

A critical challenge in this alignment process is managing the dimensional disparities between audio features and LLM embeddings. As illustrated in Fig.~\ref{fig:projector}, audio embeddings have dimensions determined by audio length $N$ and ASR encoding dimension $D_a$, while LLM embeddings are shaped by token size and embedding dimension $D_l$. The audio length $N$ typically exceeds the token count significantly due to pauses, silence periods, and other audio-specific characteristics. Directly matching these dimensions could either result in learning failure or significant information loss.

To address this dimensional mismatch, EmbedLink introduces a casted token size $T$, which serves as a fixed intermediate dimension between typical text token counts and audio lengths. After system initialization, this size remains constant to ensure training stability. For text inputs exceeding $T$ tokens, EmbedLink performs truncation; for shorter inputs, it applies padding. Meanwhile, BridgeFormer's output dimension is set to match the LLM embedding dimension $D_l$ (typically 2048 for models like Llama).

Consider an audio input processed by ASR into features $F_{\text{audio}}$ of shape $[1, N, D_a]$ and its corresponding text. The LLM embedding layer first processes the text input: if its token count exceeds the casted token size $T$, it truncates the excess; if shorter, it applies padding to reach $T$. This processed text is then converted by the LLM embedding layer into embeddings $E_{\text{text}}$ of shape $[1, T, D_l]$. Meanwhile, BridgeFormer transforms $F_{\text{audio}}$ through its architecture (as described in the previous section) to produce $E_{\text{audio}}$ with matching dimensions $[1, T, D_l]$. This alignment process can be formally expressed as:

\begin{equation}
   E_{\text{text}} = \text{LLM}_{\text{embed}}(X) \in \mathbb{R}^{1 \times T \times D_l}
\end{equation}
where $X$ represents the input text after token size adjustment.

\begin{equation}
   E_{\text{audio}} = \text{BridgeFormer}(F_{\text{audio}}) \in \mathbb{R}^{1 \times N \times D_a} \rightarrow \mathbb{R}^{1 \times T \times D_l}
\end{equation}

To optimize BridgeFormer's parameters, we employ a combined loss function that considers both direct feature matching and semantic similarity:
\begin{equation}
   \mathcal{L} = \alpha \cdot \text{MSE}(E_{\text{audio}}, E_{\text{text}}) + \beta \cdot (1 - \text{cos}(E_{\text{audio}}, E_{\text{text}}))
\end{equation}
where $\alpha$ and $\beta$ are weighting coefficients balancing the two objectives.

\subsection{Feature-based ASR encoder}


After establishing BridgeFormer and EmbedLink, selecting an appropriate ASR encoder becomes crucial for our framework. While we adopt existing ASR models rather than designing new ones, the selection must balance hardware efficiency with feature extraction capability. As discussed in Section~\ref{background:ASR}, we consider three mainstream ASR types: feature-based ASR (e.g., wav2vec2-base), transformer-based ASR (e.g., Whisper-base), and generative ASR (e.g., AudioLDM).


Our analysis combines theoretical representation space study with empirical evaluation of resource usage and training efficiency. For theoretical analysis, we examine how these ASR models map an input audio sequence $x \in \mathcal{X}$ to their respective feature spaces:

\begin{equation}
   F_{\text{audio}}^{\text{gen}} = \text{ASR}_{\text{gen}}(x) \in \mathbb{R}^{1 \times N \times D_g}
\end{equation}
where generative ASRs like AudioLDM produce variable-dimension features $D_g$ adapted to input complexity, potentially capturing the most comprehensive audio information but requiring substantial computational resources.

\begin{equation}
   F_{\text{audio}}^{\text{feat}} = \text{ASR}_{\text{feat}}(x) \in \mathbb{R}^{1 \times l(x) \times D_f}
\end{equation}
where feature-based ASRs like wav2vec2 maintain length proportional to input ($l(x)$) with fixed feature dimension $D_f$ (typically 768), offering a balanced representation of audio features.

\begin{equation}
   F_{\text{audio}}^{\text{trans}} = \text{ASR}_{\text{trans}}(x) \in \mathbb{R}^{1 \times L \times D_t}
\end{equation}
where transformer-based ASRs like Whisper use fixed dimensions ($L=1500, D_t=512$), providing consistent but potentially constrained feature representation.

\begin{figure}[t!]
  \centering
  \begin{subfigure}[b]{0.493\linewidth}
    \centering
    \includegraphics[width=\linewidth]{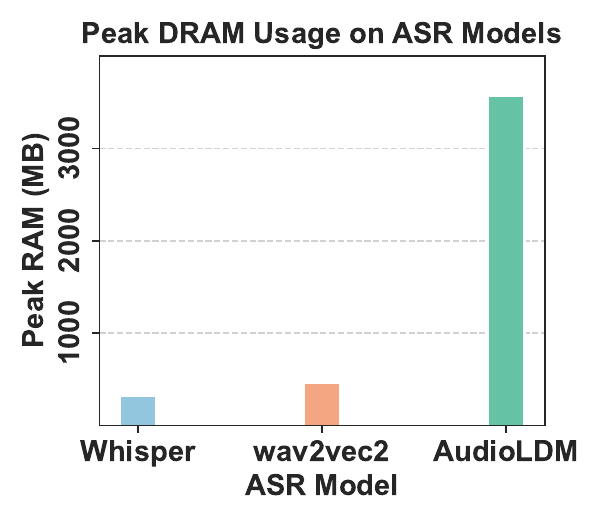}
  \end{subfigure}
  \begin{subfigure}[b]{0.493\linewidth}
    \centering
    \includegraphics[width=\linewidth]{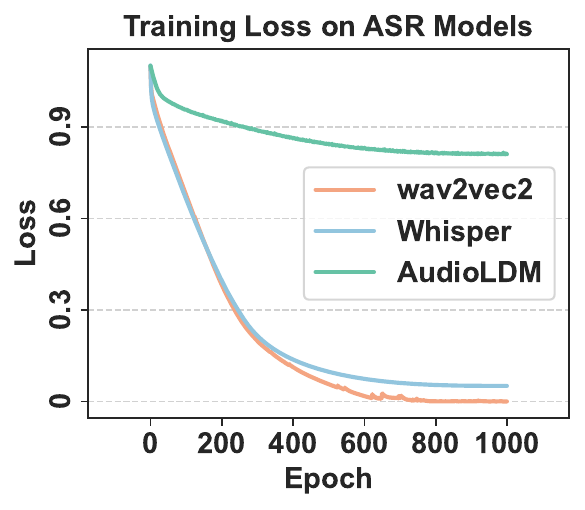}
  \end{subfigure}
  \caption{Resource efficiency comparison of three ASR models. Left: Peak RAM usage during audio feature extraction. Right: Training convergence curves using a naive MLP with features extracted from different ASR models.}
  \label{fig:preliminary_1}
\end{figure}

Our empirical evaluation, as shown in Fig.~\ref{fig:preliminary_1}, reveals two critical insights about hardware efficiency and feature quality. First, examining DRAM usage during feature extraction, we observe that generative ASRs consume significantly more memory (over 3500MB) compared to feature-based (458MB) and transformer-based (322MB) alternatives. This substantial memory footprint could severely limit the resources available for parameter learning on edge devices. Second, we evaluate feature quality through a training efficiency test using a naive MLP, tracking the cross-entropy loss over 1000 epochs. While generative ASR features show gradual, smooth convergence, feature-based ASR demonstrates remarkably rapid convergence within 400 epochs, with wav2vec2 achieving near-zero loss. Transformer-based ASR shows similar quick convergence but with slightly higher final loss values.

This comprehensive analysis leads us to select feature-based ASR, particularly wav2vec2, as the optimal encoder for our Tiny-Align framework. It offers the best balance between hardware efficiency and feature extraction capability, crucial for effective edge deployment while maintaining high-quality cross-modal alignment.

\subsection{Instruction Injection}

When the proper ASR encoder is chosen and EmbedLink has trained BridgeFormer to generate LLM-compatible embeddings, these embeddings purely represent the audio input information. However, to elicit high-quality LLM responses, instructions are often required \cite{peng2023instruction}, and these may vary across different use cases. Rather than incorporating instructions during BridgeFormer training, we design a flexible instruction injection module for inference time, allowing users to adapt instructions based on specific needs. As illustrated by the deep blue dashed lines in Fig.~\ref{fig:method}, this module implements a streamlined instruction injection process.

Given a task-specific instruction $I$, we first convert it into LLM-recognizable embeddings:
\begin{equation}
   E_{\text{inst}} = \text{LLM}_{\text{embed}}(I) \in \mathbb{R}^{1 \times N_i \times D_l}
\end{equation}
where $N_i$ represents the instruction sequence length and $D_l$ is the LLM embedding dimension. These instruction embeddings are then concatenated with the BridgeFormer output along the sequence dimension:
\begin{equation}
   E_{\text{final}} = \text{Concat}(E_{\text{inst}}, E_{\text{audio}}) \in \mathbb{R}^{1 \times (N_i+T) \times D_l}
\end{equation}
where $T$ is the casted token size used in BridgeFormer.

This concatenated embedding sequence forms a complete input for the LLM, where the instruction embeddings guide the model's interpretation of the subsequent audio-derived embeddings. This modular design allows for flexible instruction modification without requiring retraining of the alignment model, making it particularly suitable for edge deployment where adaptation to different tasks may be necessary.

\section{Experimental Evaluation}
\subsection{Experimental Setup}
\subsubsection{Datasets}
To demonstrate our Tiny-Align framework, we employ five diverse datasets from TalkBank \cite{macwhinney2007talkbank}, a comprehensive database for language research. These datasets include: (1) ADReSS-IS2020 \cite{martinc2020tackling}, containing speech recordings of Alzheimer's patients describing the Cookie Theft picture; (2) Baycrest \cite{meltzerbaycrest}, featuring structured interviews with dementia patients; (3) EllisWeismer \cite{weismer2013fast}, focusing on children with specific language impairments; (4) ENNI \cite{paradis2013discriminating}, containing narrative samples from children with language disorders; and (5) NEURAL \cite{aphasiaTalkBank}, comprising speech samples from individuals with various neurological conditions. Each dataset contains recordings from over twenty patients, with paired audio samples and corresponding transcripts. For training and evaluation, we segment these recordings into sentence-level audio-transcript pairs, resulting in approximately X training pairs and Y validation pairs. Our usage of these privacy-concerned datasets is approved by IRB \cite{grady2015institutional}.


\subsubsection{LLM Models}
We evaluate our framework across multiple state-of-the-art LLMs, including Llama-3.2-1B \cite{llama3_2}, Llama-3.2-3B \cite{llama3_2}, Phi-3.5-mini \cite{abdin2024phi}, Gemma-2-2B \cite{team2024gemma}, and StableLM-2-1.6B \cite{bellagente2024stable}. These models represent different architectures and parameter scales, allowing a comprehensive evaluation of our alignment approach. For LLM configuration, we employ maximum length padding in the tokenizer and set the temperature to 0.1 with top-k sampling (k=50) to ensure consistent and high-quality outputs.

\subsubsection{Default Experimental Setting}
For audio feature extraction, we utilize the wav2vec2-base-960h model \cite{baevski2020wav2vec} as our ASR model. It was pre-trained on 960 hours of LibriSpeech data. All audio inputs are preprocessed with a sampling rate of 16 kHz to match the model's training conditions. Our transformer-based projector employs a compact yet effective architecture with 4 attention heads, a hidden dimension of 256, and 4 transformer layers. We empirically set the casted token size to 30, which adequately captures typical sentence lengths while maintaining computational efficiency.

The training process uses the AdamW optimizer with an initial learning rate of 1e-3 and a linear decay schedule. This configuration balances training stability with convergence speed, which is particularly important for edge deployment scenarios where training resources may be limited.

The preliminary validation on a Rasberry Pi 5 can be found in section~\ref{sec:edge_validation}, followed by 
the comprehensive experiments run on a single Nvidia P100 GPU.

\subsubsection{Evaluation Methods}
We evaluate our framework from both effectiveness and efficiency perspective. For effectiveness, we employ a dual-path comparison approach: given an audio input, we compare the output from our ASR-LLM system (Output$_{A}$) with the output generated by directly feeding the ground truth transcript to the LLM (Output$_{L}$). The similarity between these outputs is measured using ROUGE scores \cite{lin2004rouge}, including ROUGE-1 and ROUGE-L, which assess word-level and longest common subsequence matches, respectively. For the ROUGE scores, higher values represent better performance of the LLM generated content.

For efficiency evaluation, we focus on convergence time as our primary metric, which measures not only the training speed but also resource utilization efficiency \cite{kim2021autofl}. Extended training periods on edge devices can lead to sustained high DRAM usage, GPU occupancy, and power consumption. We define convergence as the point where the loss stabilizes ($\Delta\text{loss} < \varepsilon$), analyzing both time-to-convergence and total resource consumption during training. This metric is particularly crucial for edge deployment, where prolonged training can strain limited device resources and impact system availability. To evaluate personalization effectiveness, we conduct user-level performance analysis by randomly sampling 20 users from each dataset and computing their average performance metrics.

\subsubsection{Baselines}
To evaluate our Tiny-Align framework, we compare it with several state-of-the-art cloud-based models. While these baselines leverage extensive computational resources and large-scale pretraining to achieve high performance, our focus remains on efficient edge deployment through training our designed projector. We select three representative baselines for comparison, where each baseline can correspond to one approach shown in Fig.~\ref{fig:preliminary_compare_method}. NExT-GPT \cite{wu2023next}, which incorporates LLM generation during ASR-LLM alignment and computes loss based on the generated content, can represent A1.X-VILA \cite{ye2024x}, which simultaneously fine-tunes both the connecting projector and the LLM during alignment, can represent A2. LanguageBind \cite{zhu2023languagebind}, which trains the ASR encoder in the edge ASR-LLM system while freezing the projector and the LLM, can represent A3. The A1, A2, A3 refer to the approaches described in Secion~\ref{sec:intro}. Compared to our method, these baselines involve more extensive parameter calculations and require substantially more computational resources.



\definecolor{darkgreen}{RGB}{0,100,0} 
\vspace{2ex}
\begin{figure}[t!]
  \centering
  \begin{subfigure}[b]{0.493\linewidth}
    \centering
    \includegraphics[width=\linewidth]{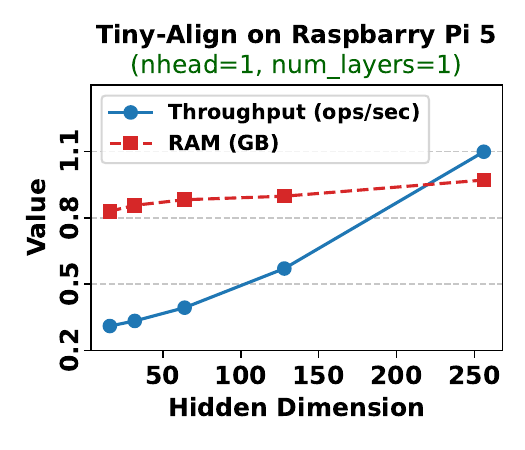}
    \captionsetup{font=small}
  \end{subfigure}
  \begin{subfigure}[b]{0.493\linewidth}
    \centering
    \includegraphics[width=\linewidth]{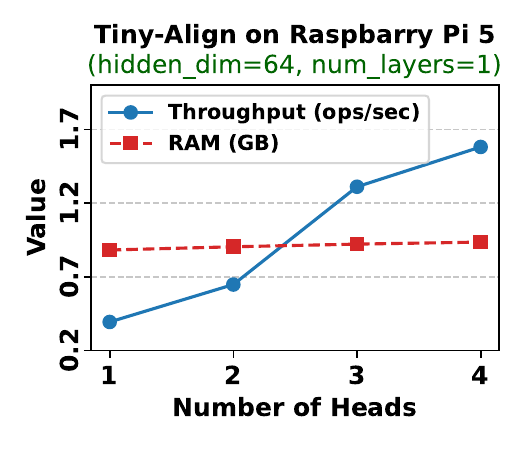}
    \captionsetup{font=small}
  \end{subfigure}
  \caption{Demonstration of the viability of Tiny-Align on a resource-constrained edge device (Raspberry Pi 5), evaluated under various configurations (x-axis) of the proposed BridgeFormer. The default settings are marked in \textcolor{darkgreen}{dark green}.}
  \label{fig:preliminary_profiling}
\end{figure}

\subsection{Results}

\subsubsection{Edge Device Implementation Validation}
\label{sec:edge_validation}
To assess whether edge device can support running the full model pipeline (i.e., ASR, projector, and LLM), 
we perform a preliminary validation on a Raspberry Pi 5. In this experiment, we evaluate two critical factors: throughput and RAM usage. 

As shown in Figure~\ref{fig:preliminary_profiling}, we explore various configurations of our BridgeFormer. The left panel illustrates the trends in throughput and memory usage across different hidden dimensions, while the right panel shows the impact of varying the number of attention heads.


\begin{figure}[ht]
    \centering

    \begin{subfigure}[b]{0.493\linewidth}
        \includegraphics[width=1\linewidth]{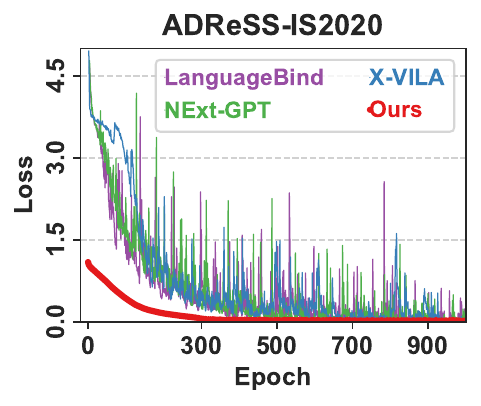}
    \end{subfigure}
        \begin{subfigure}[b]{0.493\linewidth}
        \includegraphics[width=1\linewidth]{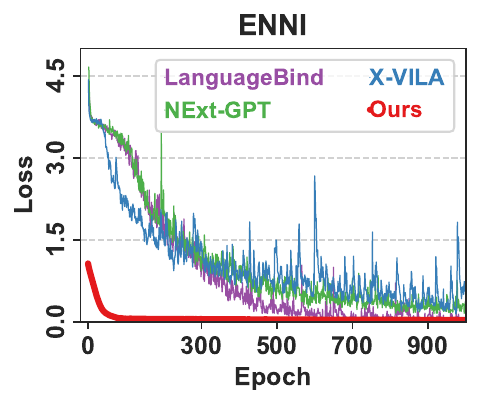}
    \end{subfigure}
    \caption{Demonstration of the loss decrease when training the wav2vec2+Gemma-2-2B based on the four methods on two datasets.}
    \label{fig:ablation_loss}
\end{figure}

\subsubsection{Comprehensive Experiments}
In this set of the experiments, we first analyze the training loss convergence of our method compared to three baselines. We evaluate an ASR-LLM system composed of wav2vec2 and Gemma-2-2B on two datasets: ADReSS-IS2020 and ENNI. Training loss is tracked over 1000 epochs. As shown in Fig.~\ref{fig:ablation_loss}, our method achieves convergence within 400 epochs on ADReSS-IS2020 and demonstrates even faster convergence (within 100 epochs) on the ENNI dataset. This rapid convergence enables efficient ASR-LLM alignment while minimizing computational resource usage and training time. Additionally, we can observe the drastic fluctuations of the loss curve for the three baselines, indicating the training under these three methods may be less stable. One thing to note is that, since the baselines are designed without considering the resource restrictions, we must limit their trainable parameters so that they can run in resource-limited environments.

\begin{figure}[t!]
    \centering
    \begin{subfigure}[b]{0.493\linewidth}
        \includegraphics[width=1\linewidth]{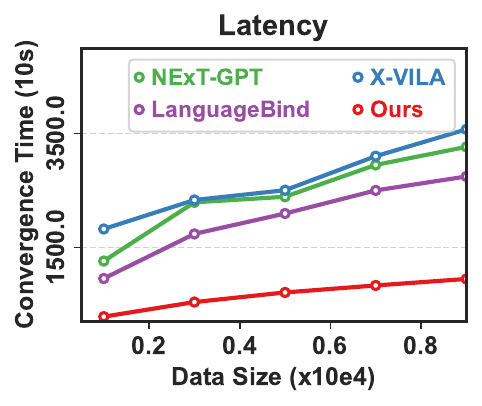}
    \end{subfigure}
        \begin{subfigure}[b]{0.493\linewidth}
        \includegraphics[width=1\linewidth]{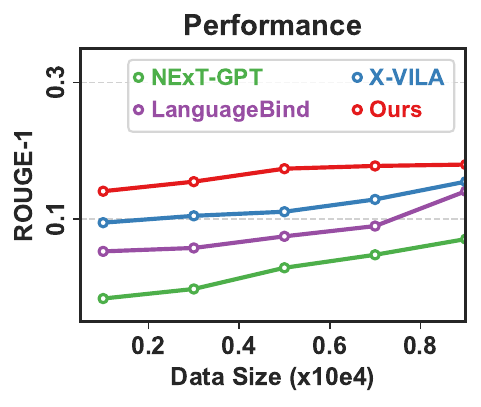}
    \end{subfigure}

    \caption{Evaluation of our method and baselines on different training data sizes given ADReSS dataset. We choose wav2vec2 and Llama-3.2-1B to formalize the ASR-LLM system. Convergence Time and ROUGE-1 score are used to measure the training time (latency) and performance.}

    \label{fig:ablation_datasize}
\end{figure}

In addition to the training loss, data size is a crucial factor in evaluating edge ASR-LLM alignment. An ideal alignment method should maintain both efficiency and effectiveness across varying data sizes, ensuring compatibility with edge devices even when users possess large volumes of data. To evaluate this aspect, we conduct experiments using the ADReSS-IS2020 dataset, creating five training data groups of different sizes. We analyze both Convergence Time and ROUGE-1 scores using our method and three baselines on an ASR-LLM system composed of wav2vec2 and Llama-3.2-1B. As shown in Fig.~\ref{fig:ablation_datasize}, our method exhibits significantly shorter convergence times and demonstrates better scaling behavior across all data sizes compared to the baselines. Furthermore, our method maintains effective ASR-LLM alignment even with limited resources, achieving comparable performance ratios between small and large data sizes. This consistent performance across different data scales demonstrates the robustness and efficiency of our method regardless of data volume.

\definecolor{cyan}{RGB}{201,151,0} 

\begin{table*}[t!]
    \caption{Evaluation of Tiny-Align (Ours), NExT-GPT (A1), X-VILA (A2), and LanguageBind (A3) on five datasets, when they are used to align the ASR and five LLM alignment. During the performance evaluation, ROUGE-1 (R-1) and ROUGE-L (R-L) are involved as the metrics and limit the training epoch up to 400. During the Latency evaluation, convergence time (C-T) is used and does not limit the training epoch to 400. }

    \centering
    {\fontsize{9pt}{18pt}\selectfont 
    \resizebox{\textwidth}{!}{
    \begin{tabular}{l l ccc ccc ccc ccc ccc}
        \toprule
        \multirow{2}{*}{\rotatebox[origin=c]{0}{Dataset}} & \multirow{2}{*}{\rotatebox[origin=c]{0}{Method}} & 
        \multicolumn{3}{c}{ASR + Llama-3.2-1B} & 
        \multicolumn{3}{c}{ASR + Llama-3.2-3B} & 
        \multicolumn{3}{c}{ASR + Gemma-2-2B} &
        \multicolumn{3}{c}{ASR + Phi-3.5-mini} &
        \multicolumn{3}{c}{ASR + StableLM-2-1.6B} \\
        & & R-1  & R-L  & C-T(10s)  & R-1  & R-L  & C-T(10s)  & R-1  & R-L  & C-T(s)  & R-1  & R-L  & C-T(10s)  & R-1  & R-L  & C-T(10s)  \\

        \midrule
        \multirow{4}{*}{\rotatebox[origin=c]{90}{ADReSS}} 
        & A1 & 0.021 & 0.025 & 293  & 0.032 & 0.027 & 293 & 0.051 & 0.049 & 2249 & 0.103 & 0.072 & 2520 & 0.030 & 0.030 & 1539 \\
        & A2 & 0.152 & 0.138 & 3107 & 0.134 & 0.128 & 3107 & 0.196 & 0.112 & 7338 & 0.213 & 0.119 & 11875 & 0.093 & 0.091 & 3670 \\
        & A3 & 0.104 & 0.096 & 1128 & 0.108 & 0.103 & 1128 & 0.185 & 0.119 & 1841 & 0.029 & 0.026 & 2093 & 0.064 & 0.060 & 1577 \\
        & Ours 
        & \cellcolor{cyan!10}{0.192} & \cellcolor{cyan!10}{0.164} & \cellcolor{cyan!10}{111} 
        & \cellcolor{cyan!10}{0.205} & \cellcolor{cyan!10}{0.193} & \cellcolor{cyan!10}{111} 
        & \cellcolor{cyan!10}{0.268} & \cellcolor{cyan!10}{0.154} & \cellcolor{cyan!10}{97} 
        & \cellcolor{cyan!10}{0.225} & \cellcolor{cyan!10}{0.121} & \cellcolor{cyan!10}{104} 
        & \cellcolor{cyan!10}{0.180} & \cellcolor{cyan!10}{0.176} & \cellcolor{cyan!10}{93} \\
        \midrule
        \multirow{4}{*}{\rotatebox[origin=c]{90}{Baycrest}}
        & A1 & 0.024 & 0.024 & 792  & 0.024 & 0.024 & 792 & 0.017 & 0.016 & 11517 & 0.050 & 0.047 & 7375 & 0.003 & 0.003 & 5055 \\
        & A2 & 0.141 & 0.104 & 949  & 0.152 & 0.109 & 949  & 0.183 & 0.114 & 4614 & 0.202 & 0.110 & 3847 & 0.095 & 0.093 & 5357 \\
        & A3 & 0.065 & 0.057 & 3015 & 0.071 & 0.067 & 3015 & 0.088 & 0.060 & 9951 & 0.012 & 0.011 & 6903 & 0.019 & 0.011 & 4918 \\
        & Ours 
        & \cellcolor{cyan!10}{0.184} & \cellcolor{cyan!10}{0.167} & \cellcolor{cyan!10}{167} 
        & \cellcolor{cyan!10}{0.197} & \cellcolor{cyan!10}{0.173} & \cellcolor{cyan!10}{167} 
        & \cellcolor{cyan!10}{0.233} & \cellcolor{cyan!10}{0.141} & \cellcolor{cyan!10}{133} 
        & \cellcolor{cyan!10}{0.205} & \cellcolor{cyan!10}{0.112} & \cellcolor{cyan!10}{183} 
        & \cellcolor{cyan!10}{0.106} & \cellcolor{cyan!10}{0.103} & \cellcolor{cyan!10}{159} \\
        \midrule
        \multirow{4}{*}{\rotatebox[origin=c]{90}{Ellis}}
        & A1 & 0.025 & 0.032 & 873 & 0.032 & 0.081 & 873 & 0.062 & 0.059 & 10593 & 0.009 & 0.009 & 5372 & 0.019 & 0.016 & 6668 \\
        & A2 & 0.187 & 0.145 & 1182 & 0.190 & 0.104 & 1182 & 0.193 & 0.107 & 3257 & 0.052 & 0.033 & 3526 & 0.044 & 0.044 & 2549 \\
        & A3 & 0.035 & 0.027 & 1563 & 0.045 & 0.051 & 1209 & 0.014 & 0.015 & 2371 & 0.022 & 0.010 & 6784 & 0.017 & 0.013 & 5366 \\
        & Ours 
        & \cellcolor{cyan!10}{0.223} & \cellcolor{cyan!10}{0.221} & \cellcolor{cyan!10}{95} 
        & \cellcolor{cyan!10}{0.223} & \cellcolor{cyan!10}{0.221} & \cellcolor{cyan!10}{95} 
        & \cellcolor{cyan!10}{0.196} & \cellcolor{cyan!10}{0.135} & \cellcolor{cyan!10}{91} 
        & \cellcolor{cyan!10}{0.146} & \cellcolor{cyan!10}{0.079} & \cellcolor{cyan!10}{136} 
        & \cellcolor{cyan!10}{0.091} & \cellcolor{cyan!10}{0.091} & \cellcolor{cyan!10}{153} \\
        \midrule
        \multirow{4}{*}{\rotatebox[origin=c]{90}{ENNI}}
        & A1 & 0.011 & 0.009 & 503 & 0.015 & 0.001 & 503 & 0.034 & 0.033 & 12530 & 0.109 & 0.076 & 13002 & 0.109 & 0.031 & 3507 \\
        & A2 & 0.109 & 0.089 & 742 & 0.157 & 0.131 & 742 & 0.191 & 0.115 & 2770 & 0.219 & 0.128 & 2165 & 0.164 & 0.164 & 1024 \\
        & A3 & 0.034 & 0.034 & 1708 & 0.034 & 0.034 & 1708 & 0.256 & 0.157 & 5407 & 0.085 & 0.073 & 5461 & 0.015 & 0.009 & 2141 \\
        & Ours 
        & \cellcolor{cyan!10}{0.124} & \cellcolor{cyan!10}{0.103} & \cellcolor{cyan!10}{91} 
        & \cellcolor{cyan!10}{0.143} & \cellcolor{cyan!10}{0.135} & \cellcolor{cyan!10}{91} 
        & \cellcolor{cyan!10}{0.244} & \cellcolor{cyan!10}{0.151} & \cellcolor{cyan!10}{82} 
        & \cellcolor{cyan!10}{0.214} & \cellcolor{cyan!10}{0.128} & \cellcolor{cyan!10}{111} 
        & \cellcolor{cyan!10}{0.303} & \cellcolor{cyan!10}{0.268} & \cellcolor{cyan!10}{74} \\
        \midrule
        \multirow{4}{*}{\rotatebox[origin=c]{90}{NEURAL}}
        & A1 & 0.023 & 0.023 & 1506 & 0.023 & 0.023 & 1506 & 0.054 & 0.048 & 17700 & 0.116 & 0.088 & 9760 & 0.009 & 0.009 & 9097 \\
        & A2 & 0.075 & 0.067 & 2187 & 0.075 & 0.067 & 2187 & 0.177 & 0.108 & 2645 & 0.157 & 0.087 & 3875 & 0.060 & 0.057 & 2356 \\
        & A3 & 0.051 & 0.048 & 1709 & 0.021 & 0.018 & 1683 & 0.034 & 0.045 & 3315 & 0.105 & 0.096 & 6149 & 0.061 & 0.055 & 6533 \\
        & Ours 
        & \cellcolor{cyan!10}{0.104} & \cellcolor{cyan!10}{0.098} & \cellcolor{cyan!10}{141} 
        & \cellcolor{cyan!10}{0.158} & \cellcolor{cyan!10}{0.109} & \cellcolor{cyan!10}{141} 
        & \cellcolor{cyan!10}{0.219} & \cellcolor{cyan!10}{0.132} & \cellcolor{cyan!10}{144} 
        & \cellcolor{cyan!10}{0.172} & \cellcolor{cyan!10}{0.093} & \cellcolor{cyan!10}{217} 
        & \cellcolor{cyan!10}{0.129} & \cellcolor{cyan!10}{0.124} & \cellcolor{cyan!10}{184} \\
        \bottomrule
        \end{tabular}
    }}

    \label{tab:model_comparison}
\end{table*}

Following our analysis of data size impact, we conduct a comprehensive evaluation of our method against the baselines. Based on our training loss analysis, we set the training epoch limit to 400 to reflect resource-constrained scenarios in ASR-LLM alignment. Given that our previous study did not show significant performance variations for different data sizes, we use a dataset of 100 samples for this comprehensive evaluation. As shown in TABLE~\ref{tab:model_comparison}, our proposed method, Tiny-Align, demonstrates superior performance across five datasets and five ASR-LLM systems, achieving significantly higher ROUGE-1 and ROUGE-L scores compared to the three baselines. Furthermore, we analyze convergence time without the 400-epoch constraint, setting a target loss of 0.05. The results, presented in the third column for each ASR-LLM system in TABLE~\ref{tab:model_comparison}, show that Tiny-Align achieves significantly faster convergence compared to the baseline methods.

\begin{figure}[ht]
    \centering
    \vspace{-2ex}
    \begin{subfigure}[b]{0.493\linewidth}
        \includegraphics[width=1\linewidth]{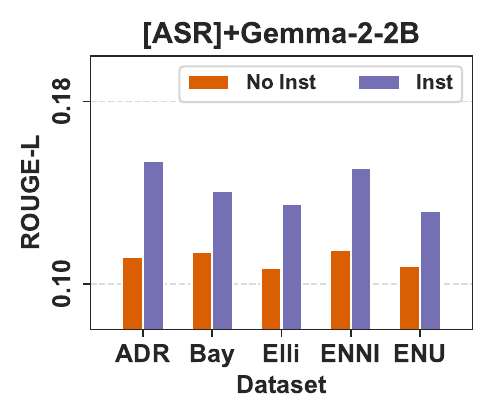}
    \end{subfigure}
        \begin{subfigure}[b]{0.493\linewidth}
        \includegraphics[width=1\linewidth]{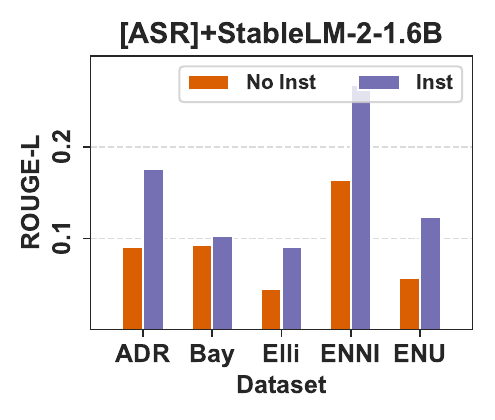}
    \end{subfigure}
    \caption{Performance comparison of instruction injection in our Tiny-Align framework across five datasets and two ASR-LLMs. The training process is unrestricted by epoch limits, allowing full convergence.}

    \label{fig:ablation_instruction_injection}
\end{figure}

Furthermore, we evaluate the instruction injection mechanism on two ASR-LLM combinations: wav2vec2+Gemma-2-2B and wav2vec2+StableLM-2-1.6B. In these experiments, we remove the epoch limit to allow complete BridgeFormer training. Instructions are then injected during the inference stage. As shown in Fig.~\ref{fig:ablation_instruction_injection}, instruction injection significantly improves LLM generation performance with the same trained BridgeFormer, indicating enhanced LLM comprehension of BridgeFormer-produced audio embeddings.




\section{Conclusion}
In this paper, we present the Tiny-Align framework, which enables edge ASR-LLM alignment. By properly designing a transformer-based projector--BridgeForme--and training it under EmbedLink, Tiny-Align can significantly lower the training cost. BridgeFormer can both effectively and efficiently build a shared embedding space, where the audio embeddings can be properly converted into LLM-compatible embeddings. With the proper ASR encoder chosen, and the boost from instruction injection, Tiny-Align results in significantly higher performance in edge ASR-LLM alignment.

\clearpage
\bibliographystyle{unsrt}
\bibliography{citations}

\begin{thebibliography}{10}

\bibitem{qin2023enabling}
Ruiyang Qin, Jun Xia, Zhenge Jia, Meng Jiang, Ahmed Abbasi, Peipei Zhou, Jingtong Hu, and Yiyu Shi.
\newblock Enabling on-device large language model personalization with self-supervised data selection and synthesis.
\newblock {\em arXiv preprint arXiv:2311.12275}, 2023.

\bibitem{sarabi2023llm}
Armin Sarabi, Tongxin Yin, and Mingyan Liu.
\newblock An llm-based framework for fingerprinting internet-connected devices.
\newblock In {\em Proceedings of the 2023 ACM on Internet Measurement Conference}, pages 478--484, 2023.

\bibitem{qin2024language}
Ruiyang Qin, Ryan Cook, Kai Yang, Ahmed Abbasi, David Dobolyi, Salman Seyedi, Emily Griner, Hyeokhyen Kwon, Robert Cotes, Zifan Jiang, et~al.
\newblock Language models for online depression detection: A review and benchmark analysis on remote interviews.
\newblock {\em ACM Transactions on Management Information Systems}, 2024.

\bibitem{hu2025combating}
Yuting Hu, Chenhui Xu, Ruiyang Qin, Dancheng Liu, Amir Nassereldine, Yiyu Shi, and Jinjun Xiong.
\newblock Combating partial perception deficit in autonomous driving with multimodal llm commonsense.
\newblock {\em arXiv preprint arXiv:2503.07020}, 2025.

\bibitem{nassereldine2024pi}
Amir Nassereldine, Dancheng Liu, Chenhui Xu, and Jinjun Xiong.
\newblock Pi-whisper: An adaptive and incremental asr framework for diverse and evolving speaker characteristics.
\newblock {\em arXiv preprint arXiv:2406.15668}, 2024.

\bibitem{yu2016automatic}
Dong Yu and Lin Deng.
\newblock {\em Automatic speech recognition}, volume~1.
\newblock Springer, 2016.

\bibitem{yang2024mala}
Guanrou Yang, Ziyang Ma, Fan Yu, Zhifu Gao, Shiliang Zhang, and Xie Chen.
\newblock Mala-asr: Multimedia-assisted llm-based asr.
\newblock {\em arXiv preprint arXiv:2406.05839}, 2024.

\bibitem{bai2024seed}
Bai et~al.
\newblock Seed-asr: Understanding diverse speech and contexts with llm-based speech recognition.
\newblock {\em arXiv preprint arXiv:2407.04675}, 2024.

\bibitem{mittal2024salsa}
Ashish Mittal, Darshan Prabhu, Sunita Sarawagi, and Preethi Jyothi.
\newblock Salsa: Speedy asr-llm synchronous aggregation.
\newblock {\em arXiv preprint arXiv:2408.16542}, 2024.

\bibitem{wu2023next}
Shengqiong Wu, Hao Fei, Leigang Qu, Wei Ji, and Tat-Seng Chua.
\newblock Next-gpt: Any-to-any multimodal llm.
\newblock {\em arXiv preprint arXiv:2309.05519}, 2023.

\bibitem{ye2024x}
Hanrong Ye, De-An Huang, Yao Lu, Zhiding Yu, Wei Ping, Andrew Tao, Jan Kautz, Song Han, Dan Xu, Pavlo Molchanov, et~al.
\newblock X-vila: Cross-modality alignment for large language model.
\newblock {\em arXiv preprint arXiv:2405.19335}, 2024.

\bibitem{zhu2023languagebind}
Bin Zhu, Bin Lin, Munan Ning, Yang Yan, Jiaxi Cui, HongFa Wang, Yatian Pang, Wenhao Jiang, Junwu Zhang, Zongwei Li, et~al.
\newblock Languagebind: Extending video-language pretraining to n-modality by language-based semantic alignment.
\newblock {\em arXiv preprint arXiv:2310.01852}, 2023.

\bibitem{radford2023robust}
Alec Radford, Jong~Wook Kim, Tao Xu, Greg Brockman, Christine McLeavey, and Ilya Sutskever.
\newblock Robust speech recognition via large-scale weak supervision.
\newblock In {\em International conference on machine learning}, pages 28492--28518. PMLR, 2023.

\bibitem{team2024gemma}
Team Gemma.
\newblock Gemma 2: Improving open language models at a practical size.
\newblock {\em arXiv preprint arXiv:2408.00118}, 2024.

\bibitem{frantar2022gptq}
Elias Frantar, Saleh Ashkboos, Torsten Hoefler, and Dan Alistarh.
\newblock Gptq: Accurate post-training quantization for generative pre-trained transformers.
\newblock {\em arXiv preprint arXiv:2210.17323}, 2022.

\bibitem{baevski2020wav2vec}
Alexei Baevski, Yuhao Zhou, Abdelrahman Mohamed, and Michael Auli.
\newblock wav2vec 2.0: A framework for self-supervised learning of speech representations.
\newblock {\em Advances in neural information processing systems}, 33:12449--12460, 2020.

\bibitem{barcovschi2023comparative}
Andrei Barcovschi, Rishabh Jain, and Peter Corcoran.
\newblock A comparative analysis between conformer-transducer, whisper, and wav2vec2 for improving the child speech recognition.
\newblock In {\em 2023 International Conference on Speech Technology and Human-Computer Dialogue (SpeD)}, pages 42--47. IEEE, 2023.

\bibitem{gulati2020conformer}
Gulati et~al.
\newblock Conformer: Convolution-augmented transformer for speech recognition.
\newblock {\em arXiv preprint arXiv:2005.08100}, 2020.

\bibitem{liu2024automatic}
Dancheng Liu, Jason Yang, Ishan Albrecht-Buehler, Helen Qin, Sophie Li, Yuting Hu, Amir Nassereldine, and Jinjun Xiong.
\newblock Automatic screening for children with speech disorder using automatic speech recognition: Opportunities and challenges.
\newblock In {\em Proceedings of the AAAI Symposium Series}, volume~4, pages 308--313, 2024.

\bibitem{yang2023diffsound}
Dongchao Yang, Jianwei Yu, Helin Wang, Wen Wang, Chao Weng, Yuexian Zou, and Dong Yu.
\newblock Diffsound: Discrete diffusion model for text-to-sound generation.
\newblock {\em IEEE/ACM Transactions on Audio, Speech, and Language Processing}, 31:1720--1733, 2023.

\bibitem{borsos2023audiolm}
Borsos et~al.
\newblock Audiolm: a language modeling approach to audio generation.
\newblock {\em IEEE/ACM transactions on audio, speech, and language processing}, 31:2523--2533, 2023.

\bibitem{majumder2024tango}
Majumder et~al.
\newblock Tango 2: Aligning diffusion-based text-to-audio generative models through direct preference optimization.
\newblock In {\em ACM Multimedia 2024}, 2024.

\bibitem{luo2023open}
Haozheng Luo, Ruiyang Qin, Chenwei Xu, Guo Ye, and Zening Luo.
\newblock Open-ended multi-modal relational reasoning for video question answering.
\newblock In {\em 2023 32nd IEEE International Conference on Robot and Human Interactive Communication (RO-MAN)}, pages 363--369. IEEE, 2023.

\bibitem{qin2024empirical}
Ruiyang Qin, Dancheng Liu, Chenhui Xu, Zheyu Yan, Zhaoxuan Tan, Zhenge Jia, Amir Nassereldine, Jiajie Li, Meng Jiang, Ahmed Abbasi, et~al.
\newblock Empirical guidelines for deploying llms onto resource-constrained edge devices.
\newblock {\em arXiv preprint arXiv:2406.03777}, 2024.

\bibitem{qin2024nvcim}
Ruiyang Qin, Pengyu Ren, Zheyu Yan, Liu Liu, Dancheng Liu, Amir Nassereldine, Jinjun Xiong, Kai Ni, Sharon Hu, and Yiyu Shi.
\newblock Nvcim-pt: An nvcim-assisted prompt tuning framework for edge llms.
\newblock {\em arXiv preprint arXiv:2411.08244}, 2024.

\bibitem{qin2024robust}
Ruiyang Qin, Zheyu Yan, Dewen Zeng, Zhenge Jia, Dancheng Liu, Jianbo Liu, Zhi Zheng, Ningyuan Cao, Kai Ni, Jinjun Xiong, et~al.
\newblock Robust implementation of retrieval-augmented generation on edge-based computing-in-memory architectures.
\newblock {\em arXiv preprint arXiv:2405.04700}, 2024.

\bibitem{qian2024linguistic}
Shengsheng Qian, Zuyi Zhou, Dizhan Xue, Bing Wang, and Changsheng Xu.
\newblock From linguistic giants to sensory maestros: A survey on cross-modal reasoning with large language models.
\newblock {\em arXiv preprint arXiv:2409.18996}, 2024.

\bibitem{xing2018enabling}
Xing et~al.
\newblock Enabling edge devices that learn from each other: Cross modal training for activity recognition.
\newblock In {\em Proceedings of the 1st International Workshop on Edge Systems, Analytics and Networking}, pages 37--42, 2018.

\bibitem{qin2024fl}
Ruiyang Qin, Yuting Hu, Zheyu Yan, Jinjun Xiong, Ahmed Abbasi, and Yiyu Shi.
\newblock Fl-nas: Towards fairness of nas for resource constrained devices via large language models.
\newblock In {\em 2024 29th Asia and South Pacific Design Automation Conference (ASP-DAC)}, pages 429--434. IEEE, 2024.

\bibitem{kramarow2024dementia}
Ellen~A Kramarow and Betzaida Tejada-Vera.
\newblock Dementia mortality among adults age 65 and older: United states, 2018-2022.
\newblock 2024.

\bibitem{vaswani2017attention}
A~Vaswani.
\newblock Attention is all you need.
\newblock {\em Advances in Neural Information Processing Systems}, 2017.

\bibitem{qin2021ibert}
Ruiyang Qin, Haozheng Luo, Zheheng Fan, and Ziang Ren.
\newblock Ibert: Idiom cloze-style reading comprehension with attention.
\newblock {\em arXiv preprint arXiv:2112.02994}, 2021.

\bibitem{peng2023instruction}
Baolin Peng, Chunyuan Li, Pengcheng He, Michel Galley, and Jianfeng Gao.
\newblock Instruction tuning with gpt-4.
\newblock {\em arXiv preprint arXiv:2304.03277}, 2023.

\bibitem{macwhinney2007talkbank}
Brian MacWhinney.
\newblock The talkbank project.
\newblock In {\em Creating and digitizing language corpora: Volume 1: Synchronic databases}, pages 163--180. Springer, 2007.

\bibitem{martinc2020tackling}
Matej Martinc and Senja Pollak.
\newblock Tackling the adress challenge: A multimodal approach to the automated recognition of alzheimer's dementia.
\newblock In {\em Interspeech}, pages 2157--2161, 2020.

\bibitem{meltzerbaycrest}
Jed Meltzer.
\newblock Baycrest pwa corpus.

\bibitem{weismer2013fast}
Susan~Ellis Weismer, Courtney~E Venker, Julia~L Evans, and Maura~Jones Moyle.
\newblock Fast mapping in late-talking toddlers.
\newblock {\em Applied Psycholinguistics}, 34(1):69--89, 2013.

\bibitem{paradis2013discriminating}
Johanne Paradis, Phyllis Schneider, and Tamara~Sorenson Duncan.
\newblock Discriminating children with language impairment among english-language learners from diverse first-language backgrounds.
\newblock 2013.

\bibitem{aphasiaTalkBank}
TalkBank.
\newblock Aphasia talkbank neural dataset.
\newblock \url{https://aphasia.talkbank.org/access/English/Aphasia/NEURAL.html}, 2024.
\newblock Accessed: 2024-11-18.

\bibitem{grady2015institutional}
Christine Grady.
\newblock Institutional review boards: Purpose and challenges.
\newblock {\em Chest}, 148(5):1148--1155, 2015.

\bibitem{llama3_2}
Meta AI.
\newblock Llama 3.2: Revolutionizing edge ai and vision with open, customizable models.
\newblock {\em arXiv}, 2024.

\bibitem{abdin2024phi}
Abdin et~al.
\newblock Phi-3 technical report: A highly capable language model locally on your phone.
\newblock {\em arXiv preprint arXiv:2404.14219}, 2024.

\bibitem{bellagente2024stable}
Marco Bellagente, Jonathan Tow, Dakota Mahan, Duy Phung, Maksym Zhuravinskyi, Reshinth Adithyan, James Baicoianu, Ben Brooks, Nathan Cooper, Ashish Datta, et~al.
\newblock Stable lm 2 1.6 b technical report.
\newblock {\em arXiv preprint arXiv:2402.17834}, 2024.

\bibitem{lin2004rouge}
Chin-Yew Lin.
\newblock Rouge: A package for automatic evaluation of summaries.
\newblock In {\em Text summarization branches out}, pages 74--81, 2004.

\bibitem{kim2021autofl}
Young~Geun Kim and Carole-Jean Wu.
\newblock Autofl: Enabling heterogeneity-aware energy efficient federated learning.
\newblock In {\em MICRO-54: 54th Annual IEEE/ACM International Symposium on Microarchitecture}, pages 183--198, 2021.

\end{thebibliography}

\end{document}